\documentclass[11pt]{article}
\usepackage{epsf}
\usepackage{amsmath}
\usepackage{graphics}
\usepackage{cite}

\renewcommand{\thefootnote}{\#\arabic{footnote}}
\begin{document}

\newcommand{\gtrsim}{ \mathop{}_{\textstyle \sim}^{\textstyle >} }
\newcommand{\lesssim}{ \mathop{}_{\textstyle \sim}^{\textstyle <} }

\newcommand{\rem}[1]{{\bf #1}}

\renewcommand{\thefootnote}{\fnsymbol{footnote}}
\setcounter{footnote}{0}
\begin{titlepage}

\def\thefootnote{\fnsymbol{footnote}}

\vskip .5in
\bigskip
\bigskip

\begin{center}
{\Large \bf $SU(3)^p$ Quiver Theories with ${\cal N} = 0$ for $p=8$ and $9$ }

\vskip .45in

{\bf Claudio Corian\`o\footnote{email:claudio.coriano@le.infin.it} and
 Paul H. Frampton\footnote{email: paul.h.frampton@gmail.com}}

{Dipartimento di Matematica e Fisica "Ennio De Giorgi"\\
 Universit\`a del Salento and INFN Lecce, 
 Via Arnesano 73100 Lecce, Italy.}

\end{center}

\vskip .4in
\begin{abstract}
\noindent
We close a gap in previous studies of nonsupersymmetric ${\cal N}=0$
quiver gauge theories from a phenomenological point of view aimed at
acquiring specific proposals for models beyond
the Standard Model (BSM). Because $SU(3)$ is the gauge
group of QCD we fix $N=3$ and vary only the $Z_p$ abelian orbifold. 
The values $1 \leq p \leq 7$
have been previously fully discussed as well as one special case, discovered by
happenstance, of $p=12$. The values
$p = 8$ and $p=9$ are discussed comprehensively in the present paper
including the electroweak mixing angle, gauge coupling unification,
spontaneous symmetry breakdown to the standard model, and
the occurrence of three quark-lepton families. Two promising quiver
node identifications are discovered for $p=8$ and three for $p=9$.
All of these merit further study as BSM candidates.
\end{abstract}

\end{titlepage}

\renewcommand{\thepage}{\arabic{page}}
\setcounter{page}{1}
\renewcommand{\thefootnote}{\#\arabic{footnote}}

\newpage

\section{Introduction}

\noindent
A possible approach to generate new models beyond the Standard Model (BSM) 
is to use non-supersymmetric gauge theories
derived from the most highly supersymmetric ${\cal N} = 4$ gauge 
theories. Such ${\cal N}=0$  theories can be systematically constructed
from ${\cal N}=4$ ones by using suitable abelian $Z_p$ orbifolding
\cite{orbifolds1,orbifolds2}.
These constructions are encoded by {\em quiver} diagrams\cite{quiver1,quiver2}, in which the $i^{th}$ node represents the $U(N)_i$ gauge symmetry and oriented arrows from the $i^{th}$ to the $j^{th}$ node represent fermions in the bifundamental $(N_i, \overline{N}_j)$ representation of the two gauge groups at nodes $i$ and $j$. Scalars are usually denoted by dashed lines connecting two nodes, in a related representation of the gauge group  $(N, \bar{N}) + (\bar{N}, N)$, if the parameters $a_i$ of the quiver theory, that we will define below, are all nonzero. If any of the 
parameters $a_i$ is zero then such scalars can be in singlet or in adjoint representations of the gauge group, but it has been shown that in such a case chiral fermions are not allowed by the theory, and as such they are of no physical interest.\\
Explicit examples with $Z_p$ orbifolding have been considered in the past for several $p$ values. 
For example, in \cite{upto7} it has been discussed a $Z_7$ model which contains
all the states of the Standard Model (SM) and in \cite{12} a $Z_{12}$ model allowing grand unification at a scale of 4 TeV. \\
The result of this construction is a gauge theory with a gauge structure of the form $SU(3)^p$ which contains a colour gauge symmetry $SU(3)_C^{n_C}$, a weak $SU(3)_W^{n_W}$ symmetry and a $SU(3)_H^{n_H}$ of hypercolour, with 
$n_H + n_W +n_C =p$. This symmetry is characterised by a single coupling $g$ above the scale of grand unification (GUT) $\mu_{GUT}$, where the $p$ factors are all independent copies of $SU(3)$, with a $Z_p$ symmetry which renders the $p$ nodes of the quiver diagram identical.
The issue whether such classically scale invariant theory may be conformal invariant at quantum level, with a vanishing $\beta$ function beyond one loop, has been matter of debate in the past, and conclusive arguments in this context are still missing\cite{Klebanov,Klebanov2}  Recent discussions of classically conformally gauge filed theories include \cite{haba1,haba2,haba3,haba4}. \\
The structure of the theory below $\mu_{GUT}$ ($\mu <\mu_{GUT}$) is of the form $SU(3)_c\times SU(3)_W \times SU(3)_H$, with a lumping of each of the $n_i$ $(n_i, i=C,W,H)$ gauge symmetries into the product of single $SU(3)$ factors of the form $SU(3)_C\times SU(3)_W\times SU(3)_H$. Each of the $SU(3)$ factors, at this scale, is the surviving diagonal subgroup of the the colour, weak and hypercolour symmetries, with couplings which are renormalized and reduced by the 
same multiplicites  $n_i$  ($g\to g_i=g/\sqrt{n_i}$). The $SU(3)^3$ symmetry of the diagonals is indeed a trinification\cite{trinification}, but with gauge couplings which are different in size and that can be unified at a far smaller scale compared to the typical $10^{15}-10^{16}$ GeV GUT scale. In ordinary trinification, the 3 couplings meet at a specific (usually very large) scale, after a large logarithmic running, which is not necessary in this case, with the result that the GUT scale can be as low as 4 TeV. \\
Above such scale, as we have already mentioned, the quiver theory is probably characterised by a quasi conformal behaviour, since the one-loop beta function vanishes, while its vanishing at two loops is not guaranteed. The appearance of double trace operators, due to the breaking of supersymmetry of the mother theory, with their non-vanishing beta-functions, has been brought up as an argument against its quantum conformal behaviour. In these theories the hierarchy is significantly ameliorated since the one loop quadratic divergences, which emerge in the Higgs sector of the SM, are absent. This is due to a precise cancellation between bosonic and fermionic contributions in the scalar 2-point function, a property which is inherited by the quiver theory from the $\mathcal{N}=4$ mother theory.\\
 In the models that we study below these features are all present and render them quite interesting from the phenomenological viewpoint. In the absence of any supersymmetric signal at the LHC, it is therefore tempting to reconsider such models in some generality, building on previous analysis and extending their classification, since they provide an alternative view to unification based on ordinary GUT's. This in an energy range which can probed at the LHC or at least at the next generation of colliders. \\
The goal of our work is to present some additional quiver theories which are consistent with the particle content of the 
SM and which have not been noticed before. In the sequence of $Z_p$ models that we consider, as we shall see,
the first with chiral fermions is $Z_4$ but
the $Z_7$ and $Z_{12}$ examples also fall into the
class we shall investigate. 

\section{General features of quiver theories}

\noindent
We consider the compactification of the type-IIB superstring
on the orbifold $AdS_5 \times S^5/\Gamma$
where $\Gamma$ is an abelian group $\Gamma = Z_p$
of order $p$ with elements ${\rm exp} \left( 2 \pi i A/p \right)$, 
$0 \le A \le (p-1)$.
The resultant quiver gauge theory has ${\cal N}$
residual supersymmetries with ${\cal N} = 2,1,0$ depending
on the details of the embedding of $\Gamma$
in the $SU(4)$ group which is the isotropy
of the $S^5$. This embedding is specified
by the four integers $A_m, 1 \le m \le 4$ with

\begin{equation}
\Sigma_m A_m = 0\,\, {\rm (mod\,\, p)}
\label{SU4}
\end{equation}
which characterize
the transformation of the components of the defining
representation of $SU(4)$.
We are here interested in the non-supersymmetric
case ${\cal N} = 0$ which occurs if and only if
all four $A_m$ are non-vanishing.
\noindent
The gauge group, ignoring $U(1)'s$, is $U(N)^p$. The fermions
are all in the bifundamental representations
\begin{equation}
\Sigma_{m=1}^{m=4}\Sigma_{j=1}^{j=p} (N_j, \bar{N}_{j + A_m})
\label{fermions}
\end{equation}
which are manifestly non-supersymmetric because no
fermions are in adjoint representations
of the gauge group.
Scalars appear in representations 
\begin{equation}
\Sigma_{i=1}^{i=3}\,\Sigma_{j=1}^{i=p}\, (N_j, \bar{N}_{j \pm a_i})
\label{scalars}
\end{equation}
in which the six integers $(a_i, -a_i)$ characterize the 
transformation of the
antisymmetric second-rank tensor representation 
of $SU(4)$. The $a_i$
are given by 
\begin{equation}
a_1 = (A_2+A_3), a_2= (A_3+A_1), a_3= (A_1+A_2).
\end{equation}

It is possible for one or more of the $a_i$ to vanish,
in which case the corresponding scalar representation
in the summation in Eq.(\ref{scalars}) is to be interpreted as an 
adjoint
representation of one particular $U(N)_j$.
One may therefore
have zero, two, four or all six of the scalar
representations, in Eq.(\ref{scalars}), in such adjoints.

Note that there is one model with all scalars in adjoints for each even
value of $p$ (see Model Nos 1,3,12). For general even $p$
the embedding is 
$A_m=(\frac{p}{2},\frac{p}{2},\frac{p}{2},\frac{p}{2})$. This series
is the complete list of ${\cal N}=0$ abelian quivers with
all scalars in adjoints.

To be of more phenomenolgical interest the model should
contain chiral fermions. This requires that the embedding
be complex: $A_m \not\equiv -A_m$ (mod p). It has been shown
that for 
the presence of chiral fermions all scalars must be in bifundamentals.

The proof of this assertion follows by assuming the contrary,
that there is at least one adjoint arising from, say, $a_1=0$. 
Therefore
$A_3=-A_2$ (mod p). But then it follows from Eq.(\ref{SU4})
that $A_1=-A_4$ (mod p). The fundamental representation of $SU(4)$
is thus real and fermions are non-chiral.

The converse also holds: If all $a_i \neq 0$ then there are chiral 
fermions.
This follows since by assumption
$A_1 \neq -A_2$, $A_1 \neq -A_3$, $A_1 \neq -A_4$. Therefore
reality of the fundamental representation would require
$A_1 \equiv -A_1$ hence, since $A_1 \neq 0$, $p$ is even 
and $A_1 \equiv \frac{p}{2}$; but then the other $A_m$
cannot combine to give only vector-like fermions.
It follows that in an ${\cal N}=0$ quiver gauge theory, chiral fermions 
are possible if and only if all scalars are in bifundamental 
representations. 

For the lowest few orders of the group $\Gamma$,
the members of the infinite
class of ${\cal N}=0$ abelian quiver gauge theories
are tabulated below. 

\begin{table}[t]
\begin{tabular}{||c||c||c|c||c|c|c||c|c||}
\hline
Model No. & p & $A_m$ & $a_i$ & scalar & scalar & chiral & Contains \\
&&&& bifunds. & adjoints & fermions? & SM fields? \\
\hline
\hline
4A & 4 & (1111) & (222) & 6 & 0 & Yes  &  No \\
\hline
\hline
5A & 5 & (1112) & (222) & 6 & 0 & Yes  &  No \\
5B & 5 & (2224) & (111) & 6 & 0 & Yes  &  No \\
\hline
\hline
6A & 6 & (1113) & (222) & 6 & 0 & Yes  &  No \\
6B & 6 & (2235) & (112) & 6 & 0 & Yes  &  No \\
6C & 6 & (1122) & (233) & 6 & 0 & Yes  &  No \\
\hline
\hline
7A & 7 & (1114) & (222) & 6 & 0 & Yes  &  No \\
7B & 7 & (1123) & (233) & 6 & 0 & Yes  &  Yes \\
7C & 7 & (1222) & (333) & 6 & 0 & Yes  &  No \\
7D & 7 & (1355) & (113) & 6 & 0 & Yes  &  Yes \\
7E & 7 & (1445) & (122) & 6 & 0 & Yes  &  Yes \\
7F & 7 & (2444) & (111) & 6 & 0 & Yes  &  No \\
\hline
\hline
\end{tabular}
\caption{List of all abelian chiral quiver models for  $p \leq 7$. }
\end{table}

We show in Table 1 the list of quiver models for  $p \leq 7$, the first is at $p=4$.
In this paper we shall 
discuss the cases $p= 8$ and $9$. We stop
at $p=9$ because we can already satisfy all of the requisite
constraints from three generations, electroweak
mixing and gauge coupling unification. More mundanely
this keeps the number of generators of the gauge group
not above 72 which is smaller than $E_6$.
In \cite{CFR} it was shown that the condition necessary for the presence
of chiral fermions, that
all the scalars must be in bifundamentals, coincides with the condition necessary
for the cancellation of one-loop quadratic divergences. This is encouraging since,
if these two conditions had been contradictory, the quiver approach would be
seriously compromised. The coincidence supports the idea that quiver gauge field theories are 
a promising and potentially fruitful future direction for BSM physics.
\subsection{ Quivers with $p > 7$} 
For $p \geq 7$, we shall keep only the chiral solutions because non-chiral
examples are of no phenomenological interest. 
We continue to number the retained models sequentially.
\noindent
Let $n_p$ be the number of inequivalent chiral quiver theories for fixed
$p$ then our search, checked by a computer program, yields the following
results: $n_2=n_3=0$,  $n_4=1$,  $n_5=2$,  $n_6=3$,  $n_7=6$ all agreeing
with the 1999 result \cite{upto7} and summarized above. Note that
for these first 12 chiral models only the $p=7$ models numbered 7B, 7D, and 7E
can have their $p$ modes labelled such that they contain the three chiral 
families of the SM. For $p=8$, we find $n_8 = 9$ with the inequivalent solutions given in Table 2 (p=8).
\begin{table}[t]
\bigskip
\begin{tabular}{||c||c||c|c||c|c|c||c|c||}
\hline
Model No. & p & $A_m$ & $a_i$ & scalar & scalar & chiral & Contains \\
&&&& bifunds. & adjoints & fermions? & SM fields? \\
\hline
\hline
8A & 8 & (1115) & (222) & 6 & 0 & Yes  &  No \\
8B & 8 & (1124) & (233) & 6 & 0 & Yes  &  Yes \\
8C & 8 & (1133) & (244) & 6 & 0 & Yes  &  No  \\
8D & 8 & (1223) & (334) & 6 & 0 & Yes  &  No \\
8E & 8 & (1366) & (114) & 6 & 0 & Yes  &  No \\
8F & 8 & (1456) & (123) & 6 & 0 & Yes  &  Yes \\
8G & 8 & (1555) & (222) & 6 & 0 & Yes  &  No \\
8H & 8 & (2222) & (444) & 6 & 0 & Yes  &  No \\
8I  & 8 & (2455) & (112) & 6 & 0 & Yes  &  No \\
\hline
\hline
\end{tabular}
\caption{All abelian chiral quiver theories with p=8}
\end{table}
\begin{table}[h]
\begin{tabular}{||c||c||c|c||c|c|c||c|c||}
\hline
Model No. & p & $A_m$ & $a_i$ & scalar & scalar & chiral & Contains \\
&&&& bifunds. & adjoints & fermions? & SM fields? \\
\hline
\hline
9A & 9 & (1116) & (222) & 6 & 0 & Yes & No \\
9B & 9 & (1125) & (233) & 6 & 0 & Yes  &  No \\
9C & 9 & (1134) & (244) & 6 & 0 & Yes  &  No \\
9D & 9 & (1224) & (334) & 6 & 0 & Yes  &  No \\
9E & 9 & (1233) & (345) & 6 & 0 & Yes  &  No \\
9F & 9 & (1377) & (114) & 6 & 0 & Yes  &  Yes \\
9G & 9 & (1467) & (124) & 6 & 0 & Yes  &  Yes \\
9H & 9 & (1557) & (133) & 6 & 0 & Yes  &  No  \\
9I & 9 & (1566) & (223) & 6 & 0 & Yes  &  No \\
9J & 9 & (2223) & (444) & 6 & 0 & Yes  &  No \\
9K & 9 & (2466) & (113))  &  6 & 0 & Yes & No \\
9L & 9 & (2556) & (122) & 6 & 0 & Yes  &  Yes  \\
9M & 9 & (3555) & (111) & 6 & 0 & Yes & No  \\
\hline
\hline
\end{tabular}
\caption{All abelian chiral quiver theories with p=9}
\end{table}

\noindent
For $p=9$, we find $n_9 = 13$ with the inequivalent solutions given in Table 3 (p=9).

\section{Model building with quiver theories}
\noindent
So far we have used only mathematics to arrive at potentially interesting chiral theories
with gauge group $SU(3)^p$  where $4 \leq p \leq 9$. More experimental data
would be very welcome to guide us beyond the SM but for the present we have to do without.

\noindent
The physics of the situation enters when we attempt to assign
the $p$ nodes to colour (C), weak (W) and hypercharge (H) preparatory to spontaneous
symmetry breaking to the SM. The labels $C$, $W$, and $H$ are for convenience with
book-keeping only. More physics constraints arise from three families, the electroweak
mixing, gauge coupling unification and the requirement of a scalar sector sufficient to
permit spontaneous symmetry breaking to the SM.

\noindent
As mentioned in the introduction, we shall use the notation
\begin{equation}
SU(3)^p \equiv SU(3)_C^{n_C} \times SU(3)_W^{n_W} \times SU(3)_H^{n_H} 
\label{qrs}
\end{equation}
for the general gauge structure of a quiver theory.
 The general understanding will be that the $n_C, n_W, n_H$ sectors will undergo spontaneously symmetry breaking to
the corresponding diagonal subgroups
\begin{equation}
SU(3)^p \longrightarrow SU(3)_C \times SU(3)_W \times SU(3)_H
\label{SSB}
\end{equation}
in which, by virtue of the choices of the diagonal subgroups, the gauge couplings
of the C, W, and H sectors are related to the original common quiver gauge coupling by
\begin{equation}
g_C = \left( \frac{g}{\sqrt{n_C}} \right) ~~~~ g_W = \left( \frac{g}{\sqrt{n_W}} \right) ~~~~ g_H = \left( \frac{g}{\sqrt{n_H}} \right).
\label{diagonal}
\end{equation}
\noindent
If we define $\alpha_i \equiv g_i^2 /(4 \pi)$ then we have from Eq.~(\ref{diagonal})
\begin{equation}
\left( \frac{\alpha_C}{\alpha_W} \right) = \left( \frac{n_W}{n_C} \right),
\label{ratioCW}
\end{equation}
which will play a role in gauge coupling unification. Notice that the other two independent ratios involving $\alpha_1$ are not necessary given the fact that the normalization of the U(1) generator is arbitrary. This will only occur if the $U(1)$ is embedded in a non abelian gauge symmetry, which is not the case here, since $U(1)_Y$ emerges both from $SU(3)_W$ and $SU(3)_H$ after the lumping of the original symmetry to diagonal.

\noindent
The electroweak mixing angle $\Theta_W$ depends on $g_W$ and on $g_Y$
where $Y$ is the weak hypercharge according to

\begin{equation}\sin^2 \Theta_W = \left( \frac{g_Y^2}{g_W^2 + g_Y^2} \right)
\label{electroweak}
\end{equation}
\noindent
From the PDG tables \cite{PDG}, the values of $\alpha_C(M_Z^2)$, $\alpha_W(M_Z^2)$
and $\alpha_Y(M_Z^2)$ at $\mu = M_Z^2= (91.19\, \textrm{GeV})^2$ are

\begin{eqnarray}
\alpha_C(M_Z^2) & = & 0.1193 \nonumber \\
\alpha_W(M_Z^2) & = & 0.03379 \nonumber \\
\alpha_Y(M_Z^2) & = & 0.010166.   \nonumber \\
\end{eqnarray}

\noindent
We shall use the RG equations
\begin{equation}
\alpha_1^{-1}(M) = \alpha_i^{-1}(\mu) - \left( \frac{b_i}{2 \pi} \right )  \ln \left( \frac{M}{\mu} \right)
\label{RG}
\end{equation}
for $ I = C, W, Y$ where the RG $\beta$-functions are, at one-loop order
\cite{EJ}
\begin{eqnarray}
b_C & = & -11 + \frac{4}{3} N_{fam} \nonumber \\
b_W & = & -\frac{22}{3} + \frac{4}{3} N_{fam} + \frac{1}{6} \nonumber \\
b_Y & = & +\frac{4}{3} N_{fam} + \frac{1}{10} \nonumber \\
\label{betas}
\end{eqnarray}
with $N_{fam} = \frac{5}{2}$ for $M \leq M_t = 173.2$ GeV and $N_{fam}=3$
for $M > M_t$.
\noindent
Using these relations, we can determine that
\begin{equation}
R(\mu) = \left( \frac{\alpha_C(\mu)}{\alpha_W(\mu)} \right)
\label{Rmu}
\end{equation}
has the value $R(\mu) = 3, 2$ for the $\mu$ values $\mu \simeq 800$ GeV, $\mu \simeq 200$ TeV, respectively,
and that
\begin{equation}
\sin^2 \Theta_W (\mu) = \left( \frac{g_Y^2(\mu)}{g_W^2(\mu) + g_Y^2(\mu)} \right)
\label{electroweakmu}
\end{equation}
has the value $\sin^2\Theta(\mu) = \frac{1}{4}$ for $\mu \simeq 3.8$ TeV. In general, for a large value of $p$, one could explore various possibilities for $R(\mu)$, linked to the ratio (\ref{ratioCW}), which would fix appropriately the unification scale $\mu_{GUT}$.

\section{Model Building for $p=8$}

\noindent
There are $n_8 = 9$ possibilities for the $A_m$ and $a_i$ listed in Table 2 (p=8) which we
may label (8A) through (8I) and analyse them in turn. We will be labelling the nodes in a quiver clockwise as nodes on a hectagon, according to their 
$C,W$ or $H$ nature and represent them in a sequence, with the edges represented by hyphens.  For $p=8$ 
\\
\\
\noindent
{\bf (8A)}  $A_m=(1115)$, $a_i=(222)$. With one color (C) node and two weak (W) node,
the node assignments allowed, when we require that there are three families and
sufficient scalars to break $SU(3)_W \times SU(3)_W$ to its diagonal subgroup, are
\noindent
\\  \\
C -W - H - W - H - H - H - H \,\,\textrm{and} C - W - H - H - H - H - H - W, \\ 
but in neither identification can the five $SU(2)_H$ be broken to a single $SU(2)_H$
subgroup because there are insufficient scalars. In order to break a product of $SU(N)$s
to their diagonal subgroup, it is necessary to have bifundamental scalars linking
all the $SU(N)$s together without dividing into subclusters.
\noindent
Thus, the (8A) quiver does not allow a 3-family SM to arise by its SSB.
Note that an oriented line C - W transforms as (3, 3*, 1) under
$SU(3)_C \times SU(3)_W \times SU(3)_H$ and the automatic anomaly cancellation
dictates that it comes only as the combination $(3, 3*, 1) + (1, 3, 3*) + (3*, 1, 3)$
which is one family. This is how to see quickly the number of families by the number
of chiral C - W links.
\\
\\
\noindent
{\bf (8B)}  $A_m=(1124)$, $a_i=(233)$. This gives the unique possible node assignment
\\ \\
\noindent
C - W - H - H - W - H - H - H
\noindent
\\
and, for this case, there are sufficient scalars for SSB to the SM.
\\
\\
\noindent
{\bf (8C)}. $A_m=(1133)$, $a_i=(334)$.
\noindent
For the W modes, a consistent node assignment would be
\\ \\
\noindent
C - W - H - W - H - H - H - H 

\noindent
but the $SU(3)_H$ groups cannot be broken to the diagonal subgroup using the scalar
bifundamentals which correspond to $a_i=(334)$.
\\
\\
\noindent
{\bf (8D)}. $A_m=(1223)$, $a_i=(334)$.

\noindent
In this case, 3-family assignments such as
\\ \\
\noindent
C - W - W - H - H - H - H - H \,\,\,   or \,\,\, C - H - W - W - H - H - H - H

\noindent
do not permit SSB of the $SU(3)_W \times SU(3)_W$ to its diagonal subgroup.
\\
\\
\noindent
{\bf (8E)}. $A_m=(1366)$, $a_i=(114)$.

\noindent
As in {\bf (8D)}, the two possible 3-family arrangements \\ \\
C - W - H - H - H - H - W - H\,\,\, and \,\,\, C - H - H - W - H - H - W - H
\\ 
\noindent
do not have the right scalars to break to $SU(3)_W$.
\\
\\
\noindent
{\bf (8F)}. $A_m=(1456)$, $a_i=(123)$.

\noindent
To obtain three familes, the $A_m$ dictate at least 3 weak W nodes whereupon
possibilities are

\noindent
\\
C - W - H - H - W - W - H - H, \,\,\,
C - W - H - H - H - W - W - H, \,\,\,  \\  \\
C - W - H - H - W - H - W - H \,\,\, and \,\,\, C - H - H - H - W - W - W - H. \\ 
For all four of these, there are sufficient scalar bifundamentals to break
the symmetry.
\\
\\
\noindent
{\bf (8G)}. $A_m=(1555)$, $a_i=(222)$.

\noindent
If we try node assignments such as
\\ \\
\noindent
C - H - H - W - H - W - H - H

\noindent
it is easy to see that there is no hope appropriately to break the $SU(3)_H$'s.
\\
\\
\noindent
{\bf (8H)}. $A_m=(2222)$, $a_i=(444)$.

\noindent
Four families are possible, and appropriate $SU(3)_W$ breaking, by assigning nodes as
\noindent
\\ \\
C - H - W - H - H -H - W - H

\noindent
but then the $SU(3)_H$ breaking is impossible for the reasons explained under {\bf (8A)}.
\\
\\
\noindent
{\bf (8I)}. $A_m=(2455)$, $a_i=(112)$.

\bigskip

\noindent
For this case, we may try either

\bigskip

\noindent
C - H - W - H - H - W - H - H\,\,\, or\,\,\, C - H - H - H - W - W - H - H.
\\ \\
\noindent
In both assignments, however, the breaking of the H's fails.

\section{Model Building for $p=9$}

\noindent
There are $n_9 = 13$ possibilities for the $A_m$ and $a_i$ listed in Table 2 (p=9) which we
may label (9A) through (9M) and analyse them in turn.

\noindent
{\bf (9A)}  $A_m=(1115)$, $a_i=(222)$. With one color (C) node and two weak (W) node,
the node assignments allowed, when we require that there are three families and
sufficient scalars to break $SU(3)_W \times SU(3)_W$ to its diagonal subgroup, are
\\
\\
\noindent
C -W - H - W - H - H - H - H - H \,\,\, or \,\,\, C - W - H - H - H - H - H - H - W
\\ \\
\noindent
but in neither case are there sufficient scalars to allow appropriate SSB of the $SU(3)_H^6$.
\\
\\
\noindent
{\bf (9B)}. $A_m=(1125)$. $a_i=(244)$.
\noindent
Three-family node identifications suggested by $A_m$ and $a_i$ are
\\ \\
\noindent
C - W - W - H - H - H - H - H - H \,\,\, and \,\,\, C - W - H - H - H - W - H - H - H\\  \\
\noindent
but the $SU(3)_W^2$ fails to break to its diagonal subgroup.

\bigskip

\noindent
{\bf (9C)} $A_m=(1134)$. $a_i=(244)$.\\ \\
\noindent
The only node identifications to try are \\ 
\noindent
C - W - H - W - H - H - H - H - H \,\,\, or \,\,\, C - W - H - H - W - H - H - H - H, \\
\noindent
but in the first the spontaneous symmetry breaking of $SU(3)_W^2$ fails, while in the
second the $SU(3)_H^6$ fails to break properly. 
\\
\\

\noindent
{\bf (9D)}. $A_m=(1224)$. $a_i=(334)$. \\ 
\noindent
Here, for three familes each of which involves a chiral C - W link we may try\\ \\
C - W - W - H - H - H - H - H - H \,\,\, or \,\,\, C - H - W - H - W -H - H - H - H \\ 
but in both cases the breaking of $SU(3)_W^2$ is impossible.\\ \\
\noindent
{\bf (9E)}. $A_m = (1233)$. $a_i = (345)$. 

\noindent
The three family structure dictates either
\noindent
C - W - H - W - H - H - H - H - H

\noindent
or

\noindent
C - H - W - W - H - H - H - H - H

\noindent
but, for both node identification choices, the $SU(3)_W^2$ symmetry breaking fails.

\bigskip

\noindent
{\bf (9F)}. $A_m = (1377)$. $a_i = (114)$.

\bigskip

\noindent
One choice C - W - H - H - H - H - H - W - H fails because of $SU(3)_W^2$ but

\bigskip

\noindent
C - H - H - W - H - H - H - W - H

\bigskip

\noindent
succeeds in that there are sufficient scalars to allow diagonal SSB of both $SU(3)_W^2$
and $SU(3)_H^6$ and thence breaking to the SM.

\bigskip

\noindent
{\bf (9G)}. $A_m=(1467)$. $a_i = (124)$.

\bigskip

\noindent
Because the four components of $A_m$ are all different, three families requires
three W nodes and, with one C node, there are four node identification choices.
One, which fails because of $SU(3)_W^3$ breaking, is \\ \\
\noindent
C - W - H - H - W - H - H - W - H. \\ \\
\noindent
The other three node identifications all work. They are\\ \\
\noindent
C - W - H - H - W - H - W - H - H, \,\,\, C - W - H - H - H - H - W - W - H\,\,\, and \,\,\, \\
C - H - H - H - W - H - W - W - H.\\ \\
{\bf (9H)}. $A_m=(1557)$. $a_i= (133)$.
\noindent
Both of the three family assignments fail in the $SU(3)_W^2$ breaking; they are\\ \\
\noindent
C - W - H - H - H - W - H - H - H\,\,\, and\,\,\, C - H - H - H - H - W - H - W - H. \\ \\ 
\noindent
{\bf (9I)}. $A_m=(1566)$. $a_i=(223)$. \\ \\
\noindent
As in {\bf (9H)}, the $SU(3)_W^2$ fails for both\\ \\
\noindent
C - W - H - H - H - H - W - H - H \,\,\, and \,\,\,C - H - H - H - H - W - W - H - H

\bigskip

\noindent
{\bf (9J)}. $A_m=(2223)$. $a_i=(444)$.\\ 
\noindent
Here it is the spontaneous symmetry breaking of $SU(3)_H^6$ which is problematic for \\ \\
\noindent
C - H - W - H - H - H - W - H - H. \\ \\

\noindent
{\bf (9K)}. $A_m= (2466)$. $a_i = (113)$. \\ \\
\noindent
$SU(3)_W^2$ symmetry breaking is impossible for both \\ \\
C - H - W - H - H - H - W - H - H \,\,\, and \,\,\, C - H - H - H - W - H - W - H - H. \\ \\

\noindent
{\bf (9L)}. $A_m=(2556)$.  $a_i=(122)$. \\ \\

\noindent
Here there is one unique node choice with enough scalars to break all the required
symmetries down to the SM gauge group. It is

\bigskip

\noindent
C - H - H - H - H - W - W - H - H

\bigskip

\noindent
{\bf (9M)}. $A_m=(3558)$. $a_i=(111)$.

\bigskip

\noindent
The diagonal subgroup of $SU(3)_H^6$ is inaccessible in \\
\noindent
C - H - H - W - H - W - H - H - H \ \ or \ \ C - H - H - H - H - W - H - H - W \\
\noindent which are the only possible 3-family choices of nodes.
\section{Discussion}

\noindent
For $p=8$, only the cases {\bf (8B)} and {\bf (8F)} allow the spontaneous symmetry breaking
to the three-family standard model, and these both have unification possible between
the gauge couplings, provided that the energy scale $\mu$ is chosen correctly.

\bigskip

\noindent
In {\bf (8B)}, the C and W embeddings require the matching condition

\begin{equation}
\frac{\alpha_W (\mu)}{\alpha_C (\mu)} = \left( \frac{1}{2} \right).
\label{8B}
\end{equation}

\bigskip
\bigskip

\noindent
In {\bf (8F)}, on the other hand, the SM embedding requires the different
condition

\bigskip

\begin{equation}
\frac{\alpha_W (\mu)}{\alpha_C (\mu)} = \left( \frac{1}{3} \right).
\label{8F}
\end{equation}

\bigskip

\noindent
Using an RGE running of the couplings $\alpha_i (\mu)$ up from the Z mass
gives the energy scales $\mu = M_{GUT} \simeq 200$ TeV
and $\mu = M_{GUT} \simeq 800$ GeV corresponding to Eqs. (\ref{8B}) and (\ref{8F}) respectively.
\noindent
For $p=9$, we have identified {\bf (9F)}, {\bf (9G)} and {\bf (9L)} as the only consistent
node identifications. The first and third require the unification implied by Eq.(\ref{8B})
while the second needs Eq.(\ref{8F}) for unification.
\noindent
These $p=8$ and $p=9$ quivers merit further study, including whether there is the possibility
of a conformal window for at least a part of the extensive energy range between $M_{GUT}$
and $M_{Planck}$.

\section{Conclusions} 
The objective of this work has been to present some additional examples of quivers which are compatible with the spectrum of the Standard Model. At the same time they involve scalars which can have VEVs 
to break the products of the SU(3)'s to the diagonal subgroups, and as such they merit further analysis. The surviving models are {\bf (8B)}, {\bf 8F)}, {\bf (9F)}, {\bf (9G)} and {\bf (9L)}. They have a type of grand unification which is quite different than the way it was envisioned long ago\cite{GUT1,GUT2}, where a single group contained the Standard Model group and that there was a desert between the weak scale and the GUT scale. The predictions of that approach were connected to proton decay and neutrino masses. In this approach, by contrast, there is no need for assumption of a desert extending over 10 or more orders of magnitude in energy. In these models the unification takes place at 800 GeV or 200 TeV which are scales within the foreseeable realm of accelerators in existence or of the next generation. They predict a wealth of new particles, including gauge bosons and further quarks and leptons. We eagerly await more data from the LHC to identify which BSM is chosen by Nature.

\noindent
\section*{Acknowledgement}

\noindent
One of us (P.H.F.) thanks INFN for hospitality and support at the University of
Salento.

\end{document}